\documentclass[submission,copyright,creativecommons]{eptcs}
\providecommand{\event}{Graphs as Models 2015} 

\usepackage{graphicx}
\usepackage{caption}
\usepackage{cite}
\usepackage{subcaption}
\usepackage{color}
\usepackage{url}
\usepackage{listings}
\usepackage{amssymb}
\usepackage{amsmath}
\usepackage{mathrsfs}

\definecolor{yellowgold}{rgb}{0.82,0.62,0.12}
\definecolor{greenforest}{rgb}{0,0.5,0}
\definecolor{bluesea}{rgb}{0.23,0.54,0.76}
\definecolor{redfire}{rgb}{0.7,0,0}
\definecolor{purpleice}{rgb}{0.65,0.51,1}
\definecolor{orangedark}{rgb}{1,0.5,0}
\definecolor{orange}{rgb}{1,0.2,0}

\newcommand{\Ra}{\Rightarrow}

\title{A Visual Analytics Approach to Compare Propagation Models in Social Networks}
\author{Jason Vallet
\institute{Univ. Bordeaux, LaBRI, UMR 5800\\F-33400 Talence, France}
\email{jason.vallet@labri.fr}
\and
H\'el\`ene Kirchner
\institute{Inria\\F-33400 Talence, France}
\email{helene.kirchner@inria.fr}
\and
Bruno Pinaud
\institute{Univ. Bordeaux, LaBRI, UMR 5800\\F-33400 Talence, France}
\email{bruno.pinaud@labri.fr}
\and
Guy Melan\c{c}on
\institute{Univ. Bordeaux, LaBRI, UMR 5800\\F-33400 Talence, France}
\email{guy.melancon@labri.fr}
}
\def\titlerunning{Visual Analytics and Network Propagation}
\def\authorrunning{J. Vallet, H. Kirchner, B. Pinaud \& G. Melan\c{c}on}

\begin{document}
\maketitle

\begin{abstract}
Numerous propagation models describing social influence in social networks can 
be found in the literature. This makes the choice of an appropriate
model in a given situation difficult. 
Selecting the most relevant model requires the ability to 
objectively compare them. This comparison can only be made at the cost of 
describing models based on a common formalism and yet independent from them. We 
propose to use graph rewriting to formally describe propagation mechanisms 
as local transformation rules applied according to a strategy. This approach 
makes sense when it is supported by a visual analytics framework dedicated to graph 
rewriting. The paper first presents our methodology to describe some 
propagation models as a graph rewriting problem. 
Then, we illustrate how our visual analytics framework allows to 
interactively manipulate models, and underline their differences based on 
measures computed on simulation traces.
\end{abstract}

\section{Introduction}

Since many years, social networks are subject to an intense research effort~\cite{Carrington:2005:MMSNA,Newman:2006:SDN,Scott:2011:SHSNA}. The study and analysis of social networks, used to represent individuals and their relations with one another, raise several questions concerning their possible evolutions.
Among these questions, the study of network propagation phenomena has 
initiated a sustained interest in the research community, offering applications 
in various domains, ranging from sociology~\cite{Granovetter:1978:TMCB,Macy:1991:CCTEC}
 to epidemiology~\cite{Hethcote:2000:MID, Dodds:2005:GMSBC, Bertuzzo:2010:SEMCE}
or even viral marketing and product placement~\cite{Domingos:2001:MNVC, Chen:2010:SIMPV}.

In this paper, we focus on propagation model analysis. The large range 
of available models and their variations make the choice of a 
particular one complicated. For instance, in~\cite{Goyal:2010:LIPSN} 
alone, the authors present three models, each with 
four variations.
Moreover, the selection cannot be solely based on simulation results because they depend on various conditions (\emph{e.g.} starting seed, unique model parameters, probability weights). 
Thus, choosing an appropriate model 
implies to effectively compare models and not only the results obtained when applying them. 

A solution is given in~\cite{Kempe:2003:MSITS} with a generalization for 
different types of propagation models. The authors allow to consider the 
models following a strict mathematical approach, thus transforming each 
algorithm in a possible solution for a common optimisation problem. On the 
contrary, we adopt an algorithmic perspective, based on graph rewriting techniques,
with the goal to consolidate an exploratory approach, based on simulation.
Propagation is usually seen as a phenomenon globally applied on a network, 
while its emerging behaviour is actually obtained from multiple local events.
Thus, most models can be represented as a set of local transformation rules, each rule describing how
an entity can influence its neighbours, and a strategy which orders and coordinates rule applications. Even if these transformations are 
depicted locally, their reiterated application allows to witness the global 
model behaviour.

In~\cite{Kejzar:2006:PICG}, the topological evolution of a social network is 
explored in a similar fashion. Starting from an existing network, the authors 
propose a set of rules to modify links between different network 
individuals, thus supporting link creation or deletion. This work intertwines with 
our approach as we both use rewriting operations to express network evolution. 
However, their article mainly focuses on network density and size 
analysis, along with the probabilistic evolutions of the generated graphs.
In the following, the considered models are used on a social network with a fixed 
topology and the rules describe how the states of nodes evolve. 

The advantages of our approach, based on a common model description, come
from the possibility to experimentally study and compare different models. Most 
of the related works concentrate on the goals achieved once the propagation 
simulation is complete (network coverage, propagation speed, \emph{etc.}), 
however
we are  more interested in determining \emph{how} such propagation occurs and 
\emph{why} those objectives are reached. The use of a common formalism allows us 
to perform these kinds of investigations. 

\begin{figure}[ht]
\centering
\includegraphics[width=\textwidth]{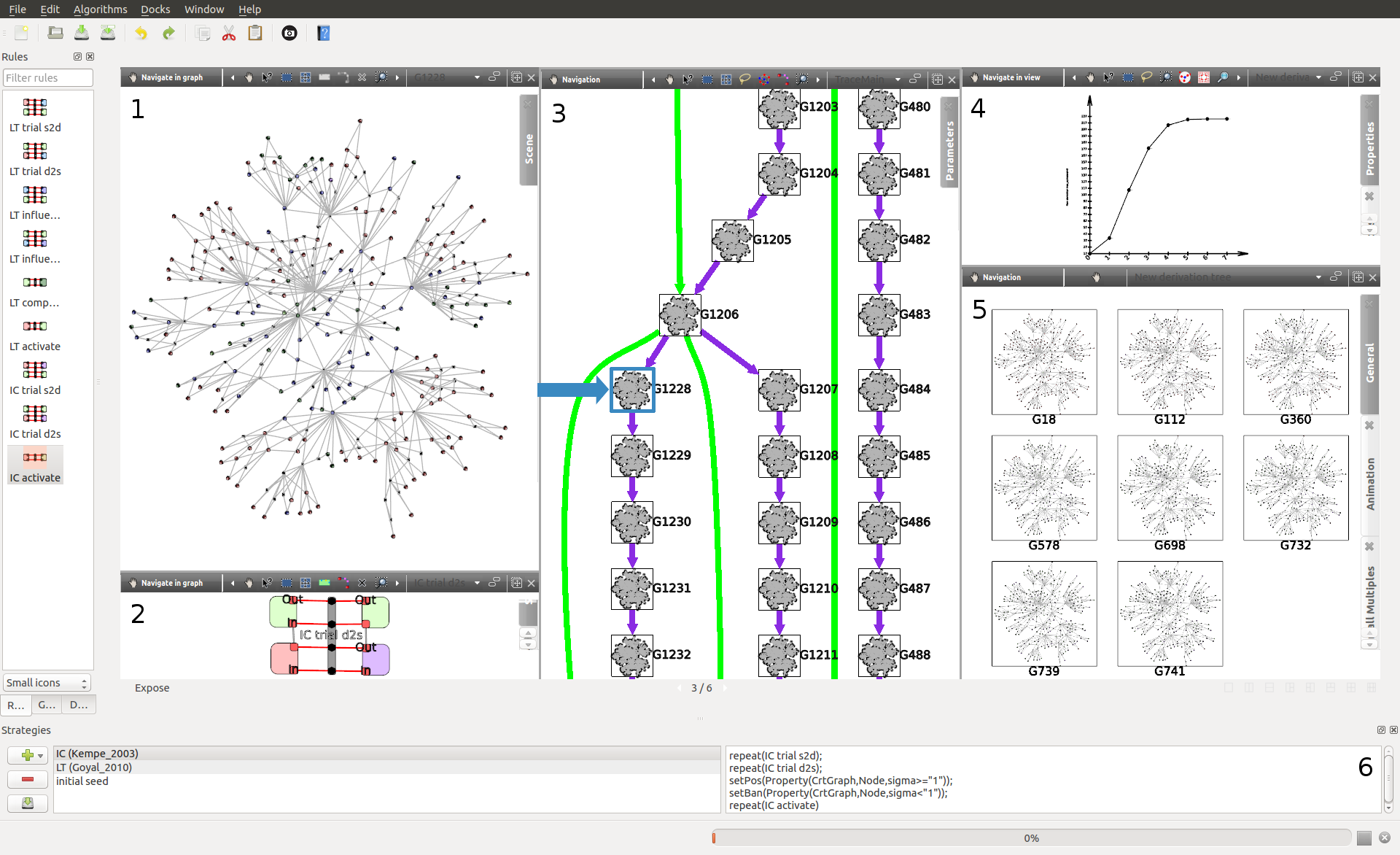}
  \caption{PORGY Interface: (1) the social network on which we apply the 
propagation; (2) rule edition; (3)  part of the derivation tree (DT), used to keep a 
complete trace of the  performed computations (\emph{e.g.} graph (1) is node $G1228$); 
(4) curve showing the number of active nodes evolution along a branch of DT; (5) another possible 
representation of a branch of DT; (6) strategy editor.}

\label{fig:overview}
\end{figure}

Furthermore, this methodology makes sense when the model study aims to be visual 
and interactive. While manipulating the model (by launching simulations, 
isolating rules, \emph{etc.}), the user is able to gradually develop a knowledge 
of the model and thus easily follow and measure its behaviour during the 
execution. For these reasons, we present a visual analytics framework --based on 
an extended version of PORGY~\cite{Pinaud:2012:PVGRE} 
(see Figure~\ref{fig:overview})-- to simultaneously build the network and the associated
rewriting rules, simulate propagations according to different strategies 
(\emph{i.e.} the models) and compare their execution traces upon various 
criteria (metrics, history, \ldots). 

This paper is organized as follows. We first present the terminology used to address social network 
propagation and describe two well-known models (Section~\ref{sect:modele}). 
Then, after a brief introduction to the graph rewriting technique used, we show how propagation
models are translated in our formalism to show the expressiveness and usability 
of graphs, rules and strategies (Section~\ref{sect:reecriture}). Finally, we show 
how the platform PORGY is used to study the propagation models and exhibit some
differences (Section~\ref{sect:visu}) and conclude by giving a few perspectives 
and future work directions.

\section{Propagation modelling in social networks}\label{sect:modele}

A social network ~\cite{Brandes:2004:AVSN} is a graph $G=(V,E)$ built from a 
set of individuals (the nodes) $V$ and a set of edges $E \subseteq V 
\times V$ linking individuals and indicating a mutual 
recognition. Two individuals, $v, w \in V$, are called neighbours when linked 
by an edge $e \in E$; we define  $N(w)$ the set of neighbours of the node $w$.
Propagation in a network can be seen as follows: when an individual performs a 
specific action (announcing an event, spreading a gossip, sharing a video clip, 
\emph{etc.}), she/he becomes \emph{active}. She/He informs his neighbours of his state 
change, giving them the possibility to become active if they perform 
the same action. Such process reiterates as the newly active neighbours share 
the information with their own neighbours. The activation can thus propagate 
from peer to peer across the whole network.

This definition is obviously simplified since each existing propagation model 
has its own specificities to be the most efficient in replicating the phenomena observed in real-world networks.
Hence some models opt for entirely probabilistic activations (\emph{e.g.}~\cite{Chen:2011:IMSNN, 
Wonyeol:2012:CATRI}) where the presence of only one active neighbour is enough to 
allow the propagation to occur. Other models use threshold values 
(\emph{e.g.}~\cite{Watts:2002:SMGCR, Kempe:2005:INDMS, Goyal:2010:LIPSN}) 
building up during the propagation. Such measures can be used to gauge the 
influence of one 
individual on his neighbours or represent his tolerance towards 
performing a given action (the more one individual is solicited, the 
more she/he becomes inclined to activate or not).

Because of the diversity across existing models and due to limited space, we limit the range of 
this paper to illustrate the feasibility of our approach on two representative 
models:~an independent cascade model \textbf{IC}~\cite{Kempe:2003:MSITS} used 
as basis in numerous others, and a linear threshold model 
\textbf{LT}~\cite{Goyal:2010:LIPSN} using a non-probabilistic activation 
principle in opposition to the independent cascade.

\paragraph{The independent cascade model (\textbf{IC})\label{sect:modele:IC}.} 
We describe a basic form as 
introduced in~\cite{Kempe:2003:MSITS}. 

\begin{itemize}
\item Let $A_0 \subset V$ be the subset of nodes initially activated at $t=0$.
\item Let $p_{v,w}$ be defined for each pair of adjacent nodes $\{v, w\}$ to 
characterize the influence probability from $v$ on $w$
($0 \leq p_{v,w} \leq 1$). $p_{v,w}$ is considered as being history independent and 
non symmetric, that is we usually have $p_{v,w} \not = p_{w,v}$.
\item A new set of nodes $A_{t+1}$ is computed from $A_t$ such as, for each $v 
\in A_{t}$, we visit the nodes $w$ adjacent to $v$ but still inactive such as 
$w \in N(v) \setminus \cup_{i = 0}^t A_i$. A given node $v$ is only offered one 
chance to influence each of its neighbours, following a probability $p_{v,w}$. 
When the adjacent node $w$ is successfully activated, it is added to $A_{t+1}$. 
\item This process continues until $A_{t+k}$ is empty ($\varnothing$) for $k 
\geq 0$.
\end{itemize}
The order used to choose the nodes $v$ and their neighbours is 
arbitrary. This model has several variations 
(\emph{e.g.}~\cite{Gomez-Rodriguez:2010:INDI, Watts:2002:SMGCR}) to allow for 
instance to simulate the propagation of diverging opinions in a social 
network~\cite{Chen:2011:IMSNN}.

\paragraph{The linear threshold model (\textbf{LT})\label{sect:modele:LT}.} This 
model behaves differently, using the neighbours' combined influence and 
threshold values to determine whether a node becomes active or stays in the 
same state.
A partial list of publications describing models using thresholds is
available in \cite{Kempe:2003:MSITS}. The model detailed below
describes the first static model defined in \cite{Goyal:2010:LIPSN}.

\begin{itemize}
\item $A_0 \subset V$ remains the subset of nodes initially activated 
at $t=0$. We also consider the probabilities $p_{v, w}$ previously introduced 
and add a \emph{threshold} value $\theta_w$ to define $w$'s resistance to its 
neighbours' influence.
\item We define $S_w$ as the set of nodes currently active and adjacent to $w$. 
For each inactive node $w$, we compute its neighbours' joint influence value 
$p_w(S_w) = 1 - \prod_{v \in S_w} (1-p_{v,w})$. 
\item $w$ becomes active when its neighbours' joint influence exceeds the
threshold value, $p_w(S) \geq \theta_w$, and $w$ is then added to $A_{t+1}$.
\item Those instructions are repeated until $A_{t+k}$ is empty ($k \geq 0$).
\end{itemize}

\section{Models translation and rewriting rules}\label{sect:reecriture}

This section describes the formalism used to express in a uniform algorithmic 
language the various models mentioned in the previous section.

Graphical formalisms are useful to easily describe complex structures in an intuitive way, like UML diagrams, proof representation, 
micro-processors design, work-flows, \emph{etc}. Extensive work exists on 
visual representation and information visualization; we refer the reader to 
\cite{Ware:2013:IVPFD} for a perception-oriented approach, and 
\cite{Ward:2010:IDVFT} for a more technical and applicative direction.

From a theoretical point of view, graph rewriting has solid logic, algebraic and 
categorical foundations~\cite{Courcelle:1990:GRALA,1997handbook1}, while from a practical
 perspective, graph transformations have many applications in specification, 
programming, and simulation tools~\cite{1997handbook2,1997handbook3}. 
Graph rewriting conveniently 
offers both semantical and operational frameworks for distributed systems and 
modelling complex systems in general. Several 
languages and tools are based on this formalism, such as PROGRES~\cite{Schurr:1999:TPALE}, 
GROOVE~\cite{Rensink:2003:GSTSS}, GrGen~\cite{Geiss:2006:GFSBG} or GP~\cite{Plump:2009:GPLGP}.

\subsection{Graph rewriting using port graphs}

Graph rewriting is a graph transformation described by
\emph{rules}, applied in a proper order specified by a 
\emph{strategy}. A rewrite rule is 
given by two graphs $L$ and 
$R$, respectively called the \emph{left-hand side} and \emph{right-hand side}. 
In a given graph $G$, the left-hand side of the rule --describing a pattern-- 
is used to identify corresponding subgraphs which should be replaced by the 
right-hand side of the rule. Different techniques can be used to describe the 
relation between $L$ and $R$, especially how the elements in the right-hand side 
replace those in the left-hand side and become connected to the 
rest of the graph. The method chosen in our solution makes use of port graphs, 
allowing the definition of such informations directly in the rewriting rules.

\subsubsection{Port graph with properties}\label{subsubsect:portgraph}

Intuitively, a port graph is a graph where nodes have explicit
connection points called {\em ports}.
A port graph $G$ is defined by a finite set $N$ of nodes $n$, a
finite set $P$ of ports $p$ (each depending from a given node of $N$) and a finite set $E$ of undirected edges. They are 
exclusively attached to ports and two ports may be 
connected by more than one edge.

Nodes, ports and edges are labelled with a set of properties.
For instance, an edge may be associated with a state (\emph{e.g.}
used or marked) and a node may have a colour, a number and a label as properties. 
Properties may be used to define
the behaviour of the modelled system and for visualization purposes
(as illustrated in examples later).

Formally, properties are pairs $(a,v)$ of an attribute $a$ and a value $v$ whose
types are described in a signature.
We write $a = v$ to mean that the attribute 
$a$ has the value $v$: for example, $active = yes$, $colour = blue$, $resist = 3$ and $size = x$ (in
the latter example we use the variable $x$ to mean that we do not care
what the value of $size$ is, just that the attribute exists).
A record $r$ is a set
 $\{(a_1,v_1),\ldots, (a_n,v_n)\}$ of properties, where each
 $a_i$ occurs only once in $r$.
All elements in $N$, $P$
and $E$ are labelled by  records defining their properties.

In this paper, to represent a social network, we use a
port graph whose nodes represent 
individuals. Among the properties attached to a
node are for instance the properties $visited$ and
$active$ which both take boolean values. Edges may also have 
properties. The model described in Section~\ref{ModelIC} uses a 
boolean property, \emph{marked}, on an edge to know if a pair of 
adjacent nodes has been visited or not.
Other properties of nodes and edges will be introduced later to
model propagation effects.

The port graph representation uses undirected edges, however, ports 
can be used to simulate edge orientation. By defining two ports on 
each node, named $In$ and $Out$, an edge becomes directed as it leaves 
a node through an $Out$ port and reaches its destination using an $In$ 
port.
In our case, we limit to one the number of edges connecting any two ports.
An edge between the $In$ port of node $A$ and the $Out$ 
port of node $B$ indicates that $A$ and $B$ are linked and can 
influence each other, regardless of the edge direction. The edge 
orientation allows us to use only one edge to store different 
properties. If $B$ influences $A$ (going from $Out$ to $In$), the 
corresponding value is quantified in the attribute $p_{o2i}$, 
expressed as a rational number; while the influence of $A$ applied on 
$B$ (from $In$ to $Out$) will be stored in $p_{i2o}$.

\subsubsection{Port graph rewrite rule}\label{subsubsect:rewriterule}

Port graphs are transformed by applying port graph rewrite rules. A {\em port 
graph rewrite rule}, noted $L \Ra R$, is itself a port graph, consisting of two 
sub-port graphs, $L$ and $R$, connected together through one special port node
$\Ra$, called {\em arrow node}.
This unique node describes how the new subgraph should be linked to the 
remaining part of the graph to avoid dangling edges~\cite{Habel:2001:DPGTR,
Corradini:1997:AAGTB} during rewriting operations.

Simple rewrite rules used in this paper are given in Figure~\ref{fig:IC_rules}.
An arrow node, coloured in grey, has a number of black ports each possessing an attribute \textit{type} 
and a set of red \emph{arrow edges}. 
These edges, connecting a port of the arrow node to ports in $L$ or $R$, are used to control the rewiring that occurs during a rewriting step.  
The ports of type \textit{bridge}, similar to those used in our example, form a bridge
between both sides of the rewrite rule, and indicate that the
corresponding port in $L$ survives the rule application.
Other types of ports correspond to more specific situations (merging, deletion, etc.) but are
not relevant in this paper. Altogether, they are meant to manage the edges 
existing between $L$ and the rest of the graph during rewriting operations.

We need to enrich this port graph structure by allowing an additional feature to 
rules: each attribute's value in the 
properties of nodes, ports and edges of the 
right-hand side may be a function of attribute's value in the 
properties of nodes, 
ports and edges of the left-hand side.
In this way, each transformation applied as a port graph rewrite rule is 
completed with the application of rule-specific functions $f$ designed to modify 
the attribute's values of each element (whether they are nodes, ports or edges). However 
this remains a local computation since the  arguments are taken from the 
properties of the 
homomorphic image of the left-hand side.
Formally, this amounts to considering values of properties which are no more 
constants in a finite domain but functions of several arguments. 
This functionality is demonstrated in Section~\ref{ModelIC} where we introduce
a node's attribute $sigma$, whose value in the right-hand side is computed from several properties originating from different elements of the left-hand side.

\subsubsection{Rewrite strategy}

To perform the appropriate graph transformations,
the rules have be to applied 
in a precise order. It is the purpose of the strategy to 
specify which rules to apply, in which order and how many times.

A strategy may also be used to depict which elements can be considered for 
the rewriting operations (matching and replacement) and which ones are 
forbidden. 
This is achieved through an additional refinement of port graphs and port graph 
rewrite rules. A \emph{located graph} $G_{Pos}^{Ban}$ consists of a port graph 
$G$ and two distinguished subgraphs $Pos$ and $Ban$ of $G$, called respectively 
the \emph{position subgraph}, or simply \emph{position}, and the \emph{banned 
subgraph}. 

In a located graph $G_{Pos}^{Ban}$, $Pos$ represents the subgraph of $G$ where
rewriting steps may take place (\emph{i.e.}, $Pos$ is the focus of the rewriting) and 
$Ban$ represents the subgraph of $G$ where rewriting steps are forbidden.  The 
intuition is that subgraphs of $G$ that overlap with $Pos$ may be rewritten only 
if they are outside $Ban$, thus restricting the application of rules
to explicit elements. $Pos$ and $Ban$ are not exclusively sets of nodes 
but may also contain edges or ports.

For instance, in a social network context, we may consider a rule
that makes a node becoming active
and limit its application to nodes whose level of incentive
is above a given threshold. This is done through a strategy $setPos$ 
that selects the adequate subgraph.

When applying a port graph rewrite rule, not only the underlying graph
$G$ but also the position and banned subgraphs may change.  A
\emph{located rewrite rule} specifies two
disjoint subgraphs, $J$ and $K$ of the right-hand side $R$, and manipulates 
these to update the position and banned subgraphs (respectively).  If
$J$ (resp.\ $K$) is not specified, $R$ (resp.\ the empty graph
$\emptyset$) is used as default.

More details on the strategy language implemented in the PORGY platform, its 
general formalisation and properties can be found in~\cite{Fernandez:2014:SPGRI}.

\subsection{Translation of propagation models}

Our main challenge concerns the translation of a
given propagation model 
into a set of rules and an adequate strategy allowing to reproduce the model 
behaviour. Graph rewriting techniques offer us to perform virtually any 
possible transformation with an appropriate rewrite rule, so one can 
always create a rule corresponding to its need. 
The real difficulty is to understand whether a finite set of rules and
strategies could describe different propagation models and their behaviours.

After a careful study of a wide range of models, we have realized the 
feasibility of such task, since
all considered models are based on either the independent 
cascade or the linear threshold models. The work 
developped in~\cite{Kempe:2003:MSITS} which proposes a framework unifying those two 
models open the way, showing that a proper generalization can bring together 
the two models, thus reducing the differences to mere variations. As we 
consider a finite number of differences between the two basic models and any 
other based on those, if one can express the \emph{IC} or the \emph{LT} model 
using a graph rewriting formalism (under the form of a strategy controlling a 
finite set of rules), any variation based on these models can also be expressed. 
Consequently, starting from the basic model translation, any extended model 
rendering can be accomplished with the introduction of a finite number of 
additional rules --to describe the differences between the currently studied 
model and the basic one-- and a few adjustments in the strategy to order  
rule applications.

We present below the translation of the IC and LT models thus providing 
instructions and showing how one should proceed to obtain similar results with 
different variations. 

\subsubsection{Model IC}
\label{ModelIC}

Using the model definition introduced in Section~\ref{sect:modele}, one can 
easily notice the two main operations to perform. The first one is the 
\emph{influence trial} where a given active node $v$ tries to influence an 
inactive neighbour $w$. The \emph{activation} of $w$, once it has been 
successfully influenced, is the second operation. Following the 
model description, these two instructions need to be applied one after the other to 
emulate the propagation process properly.

\begin{figure}
    \centering
    \begin{subfigure}[b]{0.4\textwidth}
            \centering
            \includegraphics[width=0.75\textwidth]{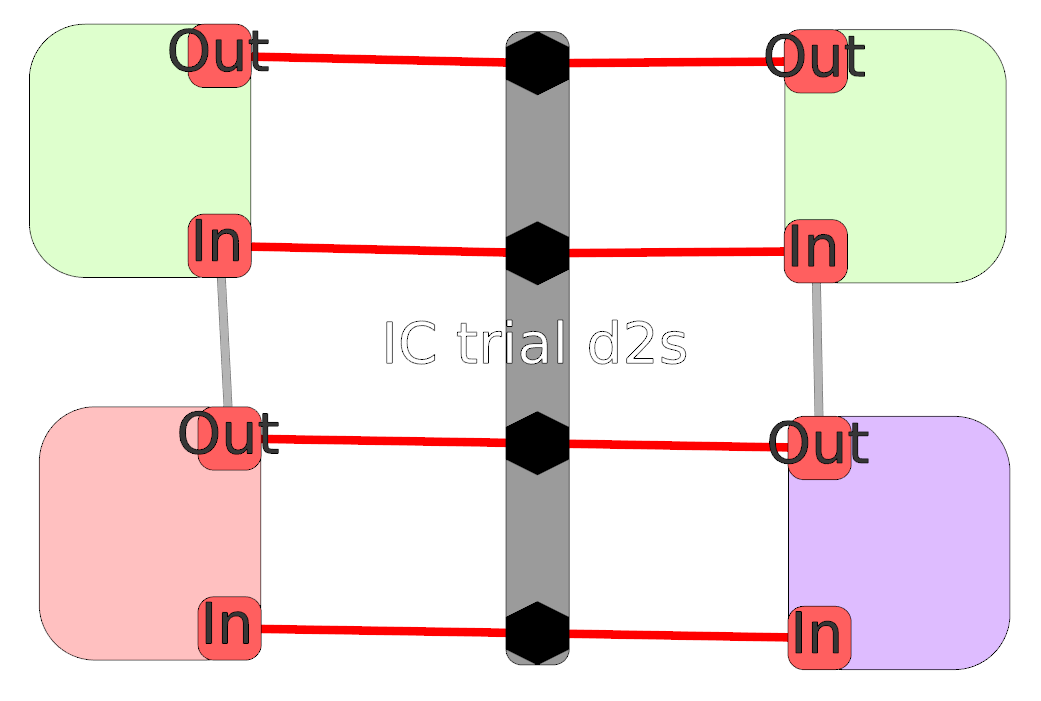}
            \caption{Influence trial from an active neighbour}
            \label{fig:IC_trial}
    \end{subfigure}
    \begin{subfigure}[b]{0.4\textwidth}
            \centering
            \includegraphics[width=0.75\textwidth]{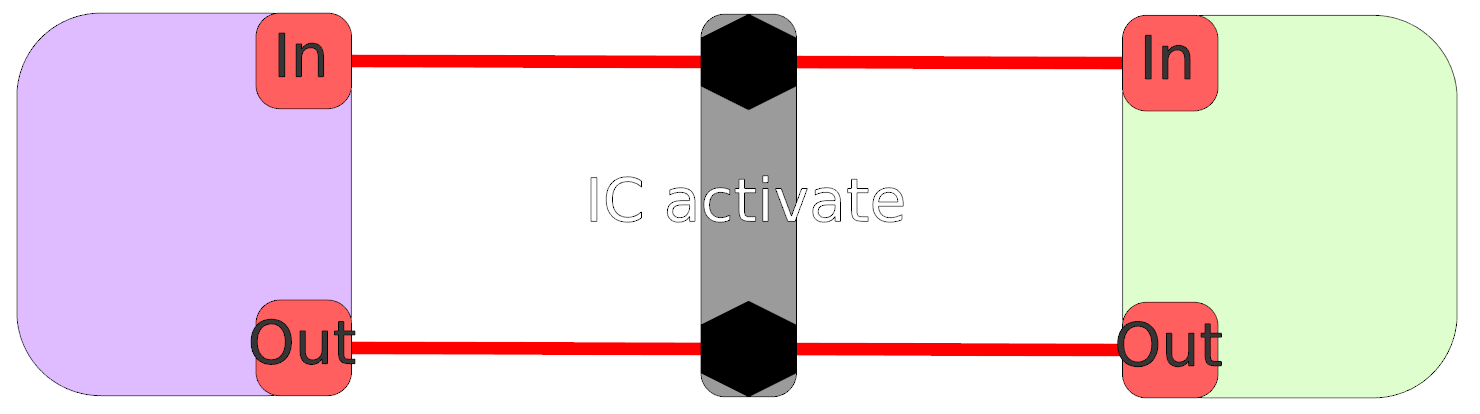}
            \caption{Visited node activation}
            \label{fig:IC_activate}
    \end{subfigure}
    \caption{Rules used to express the IC model. Active nodes 
are depicted in green and visited nodes in purple. Red nodes are in an inactive state 
(however, they may have been visited already).}
\label{fig:IC_rules}
\end{figure}

\paragraph{Rule translation}
The corresponding rules are given in Figure~\ref{fig:IC_rules}. 
Rule~\ref{fig:IC_trial} shows a pair of connected nodes in the left-hand side 
and their corresponding replacements in the right-hand side. The ports from 
each side are connected with red edges through the bridge port (as explained 
in Section~\ref{subsubsect:rewriterule}).
The node $w$, initially \emph{inactive} (in \emph{red}), is influenced 
--successfully or not-- in the left-hand side  by an \emph{active} node $v$ 
(\emph{green}). In the right-hand side, $w$ becomes \emph{purple} to
indicate visually that it 
has been \emph{visited}, while $v$ is not modified.
Finally, the edge linking the two port nodes is \emph{marked} using a boolean attribute. 
This is to limit the number of influence attempts to one for each pair of 
active/inactive neighbours. Thus, the inactive node can be visited by different 
active neighbours and be influenced several times.
Rule~\ref{fig:IC_activate} is only applied on a single node. If $w$ has been previously
sufficiently influenced by one of its neighbours, its state is changed, going 
from \emph{visited} (\emph{purple}) to \emph{active} (\emph{green}).

\paragraph{Emulating edge orientation}
For any pair of adjacent nodes $a$ and $b$, both influence probabilities 
$p_{a,b}$~(from $a$ to $b$) and $p_{b,a}$~(from $b$ to $a$) are defined as 
properties of the edge linking $a$ to $b$. The two influence probabilities 
being different, we wish to dissociate one individual from another (knowing which 
node is which) in order to use the appropriate value. As explained in 
Section~\ref{subsubsect:portgraph}, edges in port graphs are not directed, 
preventing to use edge orientation as a referential to distinguish the influence 
probabilities (one property for source to destination influence and another 
one for destination to source influence). Ports \emph{In} and \emph{Out} 
are used to emulate such edge orientation: an edge leaves by the \emph{Out} 
port and arrives through the \emph{In} port. The two properties keeping the 
influence probabilities, $p_{i2o}$~(\emph{In} to \emph{Out}) and 
$p_{o2i}$~(\emph{Out} to \emph{In}), are stored on each edge.
In Rule~\ref{fig:IC_trial}, an edge leaves an inactive node by its \emph{Out} 
port and reaches an active one through its \emph{In} port. The influence going 
from an active to an inactive node (in this case, from \emph{In} to \emph{Out}), 
we will hence use the value stored in $p_{i2o}$ (=$p_{v,w}$).
To cope with the other possibility (going from the active node to the inactive 
one in an \emph{Out} to \emph{In} direction), the rule is ``duplicated'' and 
the ports' labels are switched. 
While not presented here, this orientation is similar to the one encountered in 
the rule in Figure~\ref{fig:LT_trial}.

\paragraph{Rule inner function}
The sole execution of these rules is insufficient to perform 
a propagation. The use of the rule inner function $f$, introduced
by the end of Section~\ref{subsubsect:rewriterule}, allows us to 
perform additional
local modifications to the elements concerned by the rule 
application. For instance, in Rule~\ref{fig:IC_trial}, for each
pair
active($v$)/visited($w$): \emph{a)} we generate a random number $r \in ]0,1]$ ; 
\emph{b)} we store in a property $\sigma_{w}$ the maximum influence 
withstood by $w$ from its active neighbours until now (initialized 
to $0$) such as $\sigma_{w}=\max\left(\frac{p_{v,w}}{r}, \sigma_{w}\right)$ ; 
and \emph{c)} through a boolean property, we mark the edge linking $v$ to $w$ 
to prevent the selection of this peculiar pair configuration in the next pattern 
matching searches. This means that an active node $v$ will not be able to try to 
influence the same node $w$ over and over. 
Those local modifications, expressed through the function $f$, are applied on the 
resulting elements (from the right-hand side), using the initial element properties. 
If we define the left-hand side nodes, $v$ and $w$, and their right-hand side 
equivalents, $v'$ and $w'$, the $\sigma$ computation is expressed as:

\begin{lstlisting}[frame=single]
node(w').property("sigma") = 
	max(edge(v,w).property("p_v,w")/random(1),
		node(w).property("sigma"));
\end{lstlisting}

We affect to the property $sigma$ of the right-hand side node $w'$ the \emph{maximum} 
between the value of the property $sigma$ of the $w$ element and the result of 
the fraction $\frac{p_{v,w}}{r}$, where $p_{v,w}$ is obtained from the property 
$p\_v,w$~\footnote{$p_{v,w}$ and $p_{i2o}$ (or $p\_v,w$ and $p\_i2o$) are 
similar in Rule~\ref{fig:IC_trial}. The first notation is used to maintain 
generality.} of the edge $\{v,w\}$ and $random(X)$ returns an arbitrary value 
between $0$ and $X$. This function is locally applied after each successful application of 
the rule.
Once every active neighbour has been tried, if $w$ is 
sufficiently influenced ($\sigma_{w}>1$), it becomes itself active with 
Rule~\ref{fig:IC_activate}.

\paragraph{Strategy and rule application}
The successive applications of the rules describing the IC model have to be 
managed by a strategy. As mentioned earlier, the strategy can be for example used to 
precisely select or exclude elements as possible candidates for a rule 
application. 
The rewriting strategy used to represent the IC model is as follows:

\lstset{
numbers=left,
numbersep=5pt,
numberstyle=\tiny,
stepnumber=1
}

\begin{lstlisting}[frame=single]
repeat(IC trial d2s);
repeat(IC trial s2d);
setPos(Property(CrtGraph,Node,sigma>="1"));
repeat(IC activate)
\end{lstlisting}

The first two instructions (line 1-2) are used to call the \emph{influence trial} 
rules
\footnote{Checking each active/inactive pair of adjacent nodes, respectively the \emph{In} to \emph{Out} (Rule~\ref{fig:IC_trial}) and \emph{Out} to \emph{In} (its reciprocal) influences.} 
successively until every pair of active/inactive neighbours has been considered. 
Then (line 3), the positions in the graph where the activation should take place are 
selected as the 
nodes in the current graph whose property ``sigma'' is greater or equal to 1, 
according to our translation of the propagation model.
Line 4 executes repeatedly the activation rule at these positions.
For each rule application, the elements corresponding to the left-hand
side are chosen 
arbitrarily among the matching possibilities.
A successful and complete application of this strategy performs a round of propagation: we try to influence 
every susceptible node, then activate all the appropriately influenced ones at 
the same time. This process is reiterated to obtain successive waves, similar to 
ripples.
A more progressive behaviour can be observe by simply applying the \emph{activation} 
rule after each \emph{influence trial}. Because the choice of each pair of active/inactive nodes is arbitrary, 
such activation process can be seen as a ``random'' propagation. However, due 
to limited space, this kind of propagation evolution is not addressed in this paper.

\subsubsection{Model LT}
\label{ModelIT}

The second model seems more complicated but the approach is very similar. Here again, two different operations are used to perform the 
propagation: for each inactive node, we compute the joint influence of its 
active neighbours, then, if the influence exceeds the threshold 
value, the node becomes active.

Before presenting the corresponding rules, a few precisions need to 
be highlighted concerning this model described in~\cite{Goyal:2010:LIPSN}. The paper 
presents, among others, a static propagation model with several variations. We 
need to define the probability of $v$ influencing $w$ ($p_{v,w}$), however, such 
probability of influence from one 
individual to another may change from time to time 
(additional details are given in~\cite{Goyal:2010:LIPSN}). Because the activation of a specific
 node $w$ is dependent of the probabilities emanating from each of its neighbours, 
we need to keep their joint influence  $p_{w}(S_w)$ updated.

Such correction can be performed using the two operations $p_{w}(S_w \setminus 
\{u\})$ (Equation~\ref{eq:difference}), suppressing the influence of $u$ on $w$ among 
its neighbours ($S_w$), and $p_{w}(S_w \cup \{u'\})$ (Equation~\ref{eq:union}), adding 
the influence of $u'$ among the other adjacent nodes of $w$. 
\begin{equation} \label{eq:difference}
p_{w}(S_w \setminus \{u\}) = \frac{p_{w}(S_w) - p_{u,w}}{1 - p_{u,w}}
\end{equation}
\begin{equation} \label{eq:union}
p_{w}(S_w \cup \{u'\}) = p_{w}(S_w) + (1 - p_{w}(S_w))*p_{u',w}
\end{equation}
An improvement, introduced in~\cite{Goyal:2010:LIPSN}, 
gives the opportunity to 
update the joint influence in a single operation. The two manipulations on the 
set $S_w$ --deleting the previous influence probability value with 
Equation~\ref{eq:difference} and adding the new one using Equation~\ref{eq:union}-- can be 
perform simultaneously: 
$p'_{w}(S_w) = p_{w}(\{S_w \setminus \{u\}\} \cup \{u'\}) = p_{w}(\{S_w \cup 
\{u'\}\} \setminus \{u\})$.

\begin{figure}
    \centering
    \begin{subfigure}[b]{0.4\textwidth}
            \centering
            \includegraphics[width=0.75\textwidth]{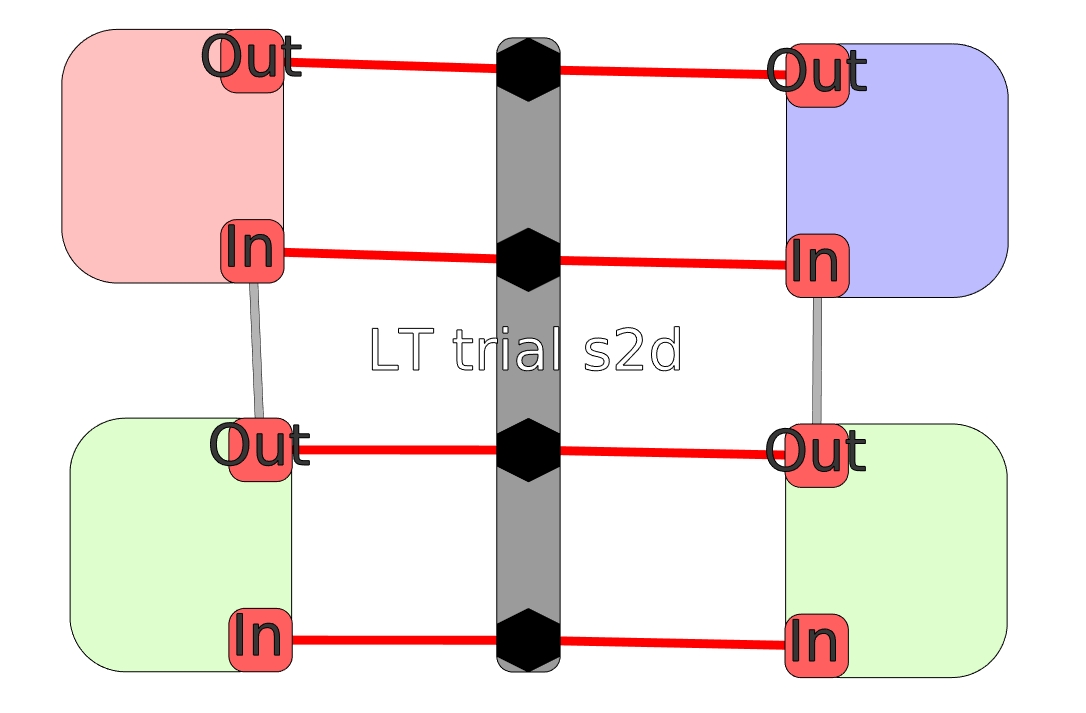}
            \caption{Joint influence computation from an active neighbour}
            \label{fig:LT_trial}
    \end{subfigure}
    \begin{subfigure}[b]{0.4\textwidth}
            \centering
            \includegraphics[width=0.75\textwidth]{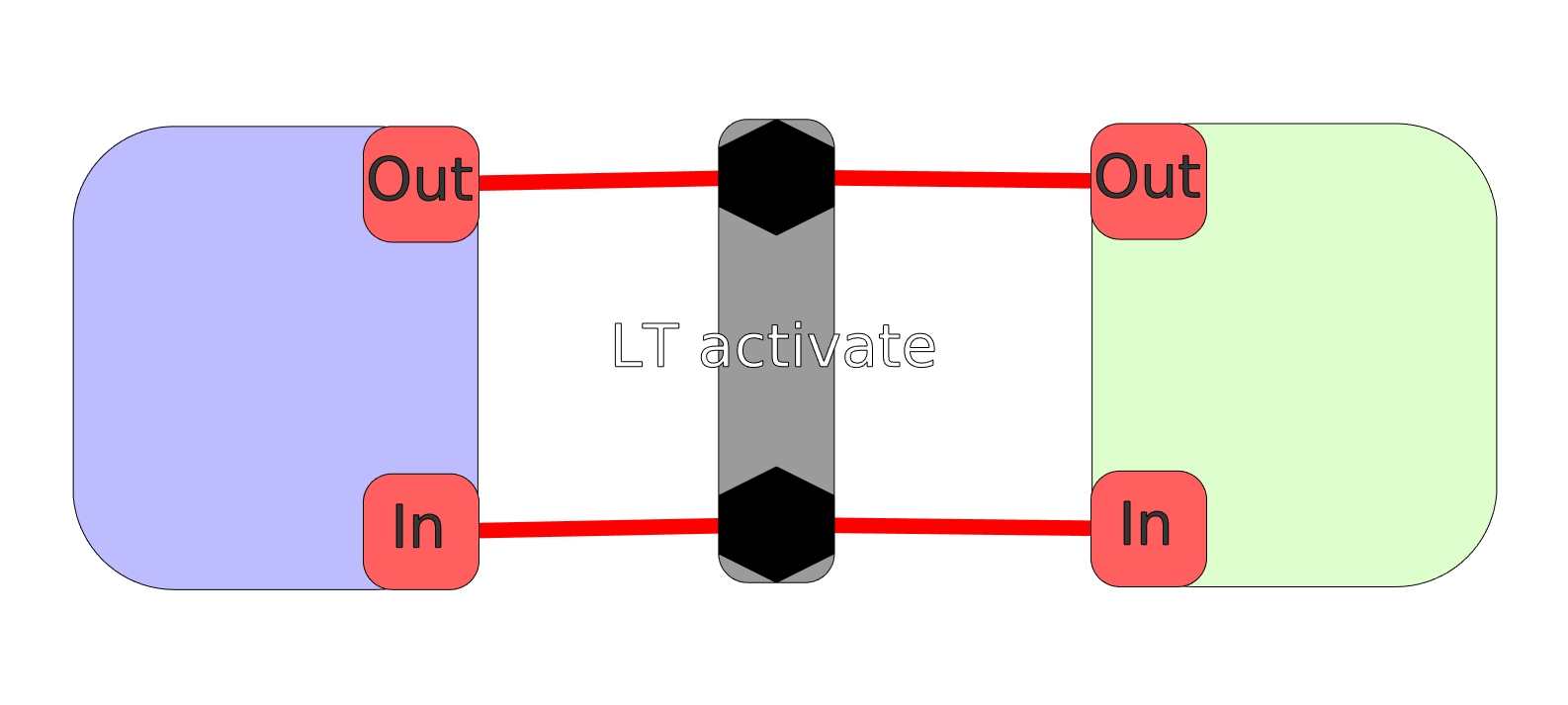}
            \caption{Influenced node activation}
            \label{fig:LT_activate}
    \end{subfigure}
    \caption{Rules used to express the linear threshold model. Colours conserve 
the same meaning as previously: active nodes are green, visited nodes are purple 
and red nodes are inactive (but may have been already visited).}
\label{fig:LT_rules}
\end{figure}

\paragraph{Rules and strategy}
The rules created from this model are quite similar to those introduced before. 
The first one (Figure~\ref{fig:LT_trial}) is applied on a pair of active/inactive 
nodes (respectively green and red). It comes in two variations depending of the 
\emph{In/Out} port used.
As we consider the possibility of an updated influence probability from $v$ to 
$w$ ($p_{v,w}$), when trying to activate, $w$ has to refresh the joint influence 
($p_{w}(S)$) of its neighbours. The inner function performed when the rule is 
applied computes $p'_{w}(S) = p_{w}(\{S \setminus \{u\}\} \cup \{u'\})$, 
 where $u$ is the neighbour using its previous influence probability and $u'$ is
the node with the updated value. For each application, a $\sigma_w$ 
value is also calculated as follows: $\sigma_w =\frac{p_{w}(S)}{\theta_w}$, so it 
becomes greater or equal to $1$ when $w$ is ready to activate ($\theta_w$
 --stored in a property-- being 
the threshold value that the joint influence $p_w(S)$ must reach for $w$ to activate).
The second Rule~\ref{fig:LT_activate} is \emph{per se} identical to the 
\emph{activate} rule shown in Figure~\ref{fig:IC_activate}. A successfully 
influenced node is simply activated; the strategy alone is used to delimit the 
nodes eligible for activation.

Like the IC model, the first two rules are applied successively 
on the graph to evaluate every active/inactive pair. The ``sigma'' property is 
once more used to determine upon which nodes the activation must be performed. 
Eventually, the activation rule is applied as many times as needed:

\begin{lstlisting}[frame=single]
repeat(LT trial s2d);
repeat(LT trial d2s);
setPos(Property(CrtGraph,Node,sigma>="1"));
repeat(LT activate)
\end{lstlisting}

Alike the previous model, successive applications of the strategy creates rounds of propagations. If one wishes to obtain a more progressive and irregular behaviour, the \emph{activation} rule must be applied after each \emph{influence trial}.

\section{Visual analytics and model comparison}
\label{sect:visu}

We detail in this section how the 
PORGY platform  is used to 
compare two series of propagations from the independent cascade and linear 
threshold models. To illustrate our method, we create 
a random social network using the graph generation model introduced in~\cite{Wang:2006:RPSFN}. The resulting graph contains 300 nodes 
(chosen as a parameter) and 597 edges (defined by the generator)\footnote{In this 
example, a small network has been preferred to improve the visual aspect of the 
screenshots but similar analysis can be performed on wider graphs.}. The 
starting conditions are similar for the two propagation models: same set of 
initially active nodes and same starting values for the different properties 
needed by the models (\emph{e.g.} influence probability distribution, threshold 
value, \ldots). We do not aim here at showing if a given model is better or more 
realistic than another one, as such task needs numerous simulations to 
compute the average results on the probabilistic models and extensive
comparisons with real-world cases.
Moreover, the considered propagation models have already been
validated in that respect.
Instead, we want to 
understand how models evolve and behave\footnote{Every element (software, 
data, \ldots) needed to reproduce our results are available at 
\url{http://tulip.labri.fr/TulipDrupal/?q=porgy}.}.

\subsection{Network evolution and history}

Upon each rule execution, 
an intermediate state of the original graph is created and stored in the 
propagation trace, thus keeping track of the changes applied to each node and 
edge in the graph, for instance nodes going from \emph{unvisited} to 
\emph{visited} and possibly \emph{active}. The history of those modifications, 
or \emph{rewriting steps}, can be used to study and compare two states of 
the graph at a given time or to rebuild and follow the path 
of node activation.

\begin{figure}
    \centering
    \begin{subfigure}[b]{0.32\textwidth}
            \includegraphics[width=\textwidth]{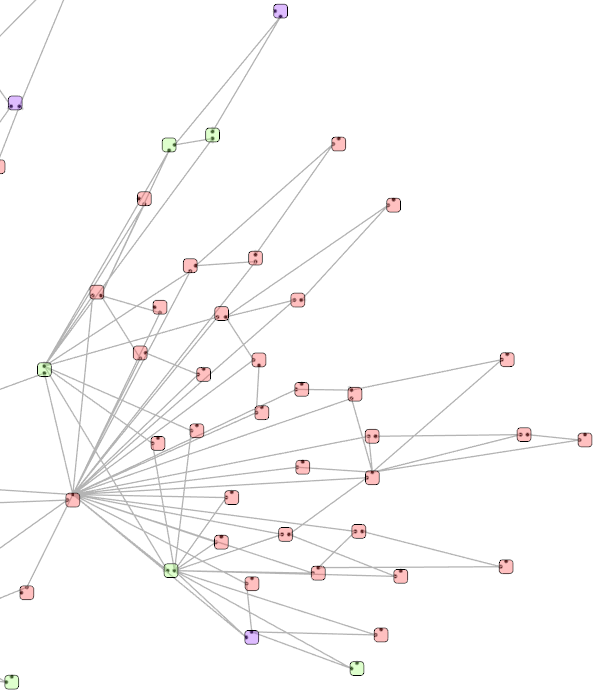}
            \caption{$t=1$}
            \label{Wang_propagation_G112}
    \end{subfigure}
    \vline
    \begin{subfigure}[b]{0.32\textwidth}
            \includegraphics[width=\textwidth]{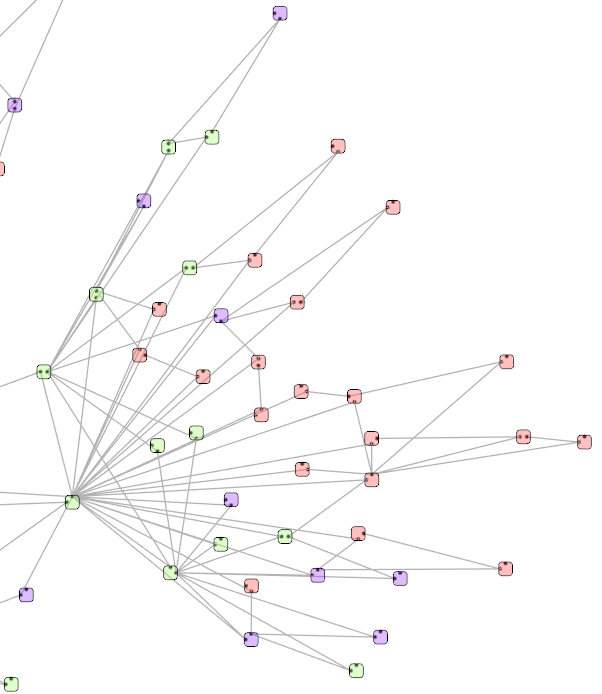}
            \caption{$t=2$}
            \label{Wang_propagation_G360}
    \end{subfigure}
    \vline
    \begin{subfigure}[b]{0.32\textwidth}
            \includegraphics[width=\textwidth]{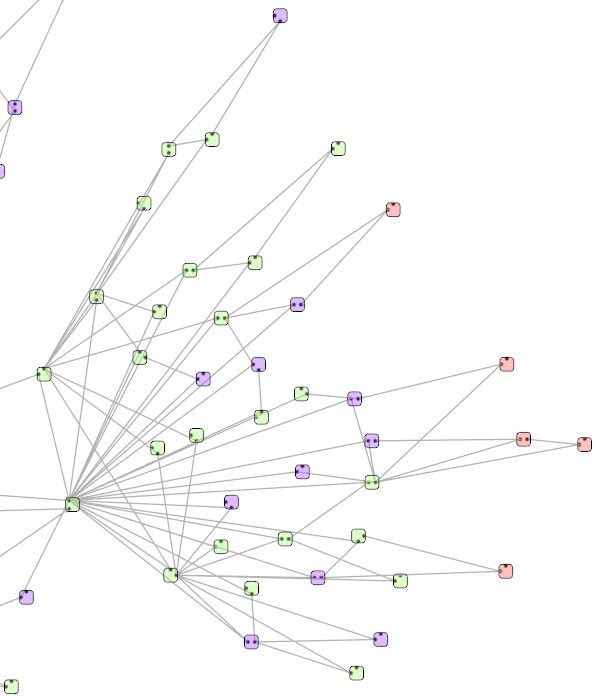}
            \caption{$t=3$}
            \label{Wang_propagation_G578}
    \end{subfigure}
    \caption{Sub-community extracted from the graph showing the propagation 
evolution at different time steps. Two nodes are activated at $t=0$ (not showed here) and six are 
 only visited at the end of the propagation (at $t=7$). (Legend: green = active, purple = 
visited, red = unvisited and inactive)}
\label{fig:Wang_propagation}
\end{figure}

A derivation tree (see Figure~\ref{fig:overview}) is created and maintained to 
provide these information.
To improve readability, several rule applications are grouped together
according to the strategy execution and considered as one \emph{propagation step}.
The successive states visualization allows us to witness this progression. 
Figure~\ref{fig:Wang_propagation} presents several representations of 
the same subgraph, showing node states evolution. The 
different timestamps $t$ indicate the successive strategy applications (an 
unvisited node at time $t$ can thus become activated at $t+1$). 

The derivation tree allows us to immediately identify the model which needs the less 
propagation steps (\emph{i.e.} less strategy application steps) before concluding the 
propagation, since it corresponds to the shorter branch (Figure~\ref{fig:Comparison}). We thus 
obtain a first complexity approximation concerning algorithms execution.

\subsection{Statistics}

\begin{figure}
    \centering
    \includegraphics[width=\textwidth,keepaspectratio]{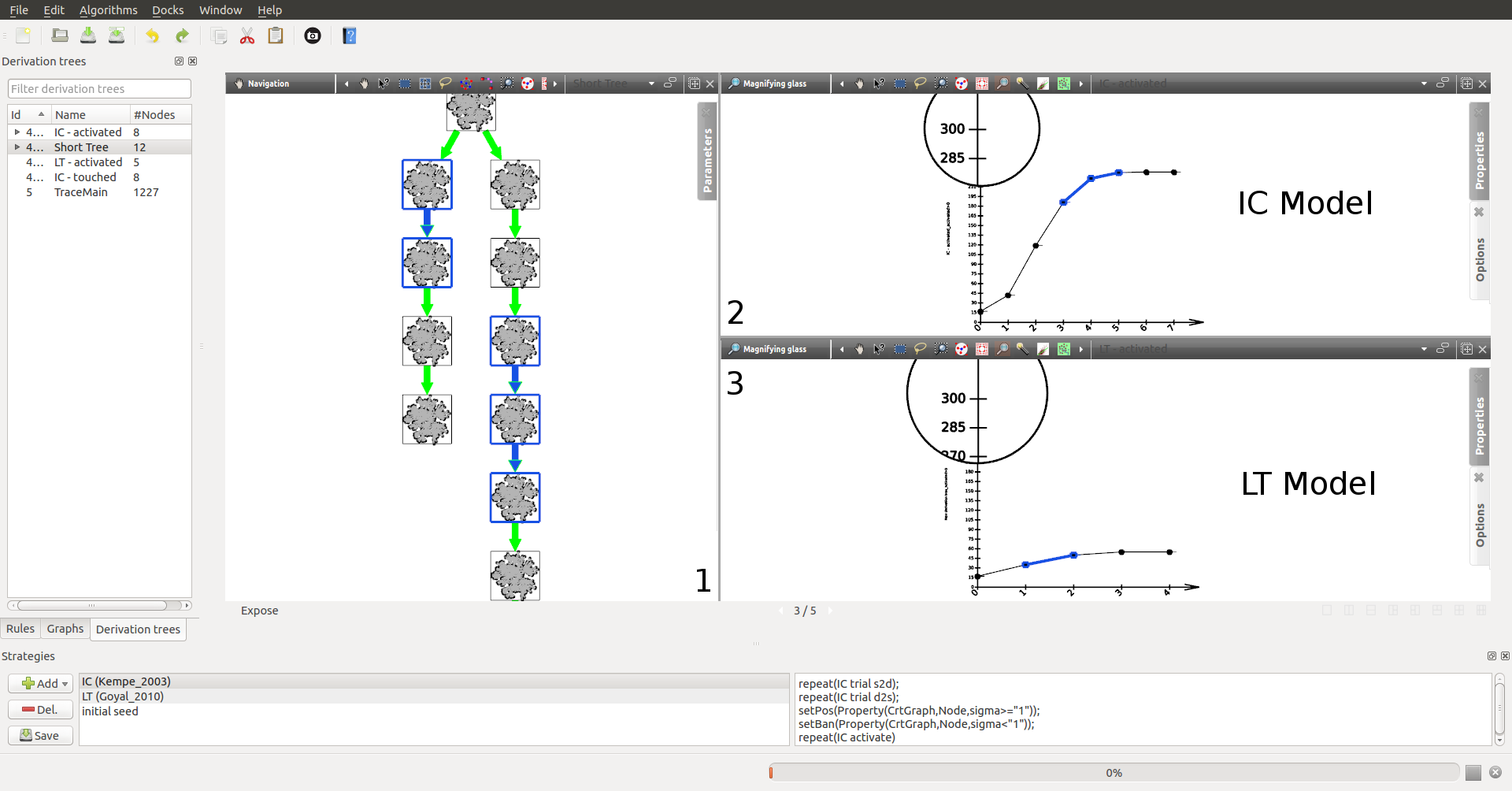}
    \caption{PORGY interface: (1) portion of a simplified derivation tree 
(filtered to show only the results obtained after a whole strategy 
application) ; (2) and (3)  \emph{Propagation speed} evolution (number of active 
nodes) according to the depth in the derivation tree branch for the independent 
cascade (IC) and linear threshold (LT) models.}
\label{fig:Comparison}
\end{figure}

Linking the derivation tree depth (equivalent to the number of 
propagation steps along a branch) with 
other measures allows us to consider the evolution of several 
parameters all along the propagation. We can, for instance, use the notion of 
\emph{propagation speed}, a value obtained by examining the number of active 
nodes at each step and evaluate their evolution. Figure~\ref{fig:Comparison} 
presents such measures for an independent cascade model ((2), top right panel) 
and a linear threshold model ((3), bottom right panel) executions. 
The x-axis displays the depth (maxed out at $7$ for (2) and at $4$ for (3)) and the y-axis  the number of active nodes (scaling from $0$ to $300$ in both curves). 
We can notice that the cascade 
model has been able to activate approximately $80\%$ of its nodes versus $18\%$ 
for the threshold model (for a similar number of steps), thus showing the high 
performance impact on different models when using identical values for the 
influence probability initialization and the same starting set of active nodes. We 
have used a magnifying glass (a functionality available in PORGY) on the top of 
both axes to improve the scale readability. We can also note the number of 
activated nodes after the first strategy application. The values measured on 
both scatterplots are initially close, the gap between the two models 
getting wider with each new application.

Moreover, time-dependent measures can be generalized to other propagation 
properties. The notion of \emph{visited nodes}, introduced earlier, can bear 
some interest when the propagation subject is only a message which needs to be 
seen and not merely disseminated or spread by willing 
individuals.
Such 
\emph{acknowledgement speed} of the propagation content can be determined like 
the propagation speed. Moreover, considering those two measures, we can 
introduce a third one defining the \emph{propagation efficiency}, computed as 
the ratio of the number of active nodes at time $t$ against those influenced/visited 
at $t-1$. Several additional metrics can be computed depending of the wished 
analysis as PORGY offers us to choose any of the existing element property and 
traces its evolution through the derivation tree. 

\subsection{Tracing propagation behaviour}

One can also be interested in observing the state of the social network when the number 
of activated nodes reaches a given value. Because the different views in the 
software are linked together, any operation in one view impacts the other views. A 
node/edge selection (highlighted in blue in Figure~\ref{fig:Comparison}) in the
 derivation tree implies the same in the scatterplot and \emph{vice versa}. A 
similar kind of operation in one of the graph states will echo through all the 
states containing the same elements, thus highlighting the selected nodes/edges 
everywhere they are. 
Such behaviour is available thanks to the inner datastructure of PORGY, leaving 
the nodes unmodified as long as they are not subject to a rule application. 
Consequently, selecting an element of interest in one of the intermediate graphs 
allows us to follow and determine in which state this element is altered. Such 
operation is especially useful when working with the complete derivation tree 
(showing the details of every rule application).

Using the final propagation state, one can witness whether or not a node has 
finally been able to perform the action (activated). Identifying the influenced but inactive 
node and the unvisited elements may also help in finding the nodes 
responsible for limiting the propagation. Once the wrongdoers have been found, 
specific treatments can target them in order to improve their condition 
(\emph{i.e.} lowering their resistance or rising their neighbours' influence). 
A new simulation can then be launch to check if the modifications have been 
successful.

\section{Conclusion and future work}

We have presented a graph rewriting based formalism seen as a common language to 
express network propagation models. When the propagation does not imply 
topological modifications in the graph, models can be described as a reduced 
set of transformation rules used to 
express network state transitions and a strategy to manage their application. 

In order to demonstrate the generality of our approach, we plan to use more propagation models  and multiply the 
number of simulations.
Simulating propagation using our method may be performed on large
networks as well. Nevertheless, this raises additional challenges
concerning the scalability of our approach and the possible visual
analysis. The search for left-hand side 
matching (graph-subgraph
isomorphism) is a demanding operation in big graphs even when, because of the limited number
of elements in the rules left-hand side, the time complexity is polynomial. 

Looking for propagation minimization or maximization on large networks by visual means may also be impeded by the combinatorial explosion of possible states thus limiting such operations.
In general, visualization of large graphs is a complicated issue. Resorting to node-link views, such as the ones presented in this paper, usually gives poor results when the number of nodes or edges exceeds a few thousands. Alternative solutions, such as matrix-based hybrid~\cite{Henry:2007:NHVSN,Rufiange:2012:THVCG} or pixel-oriented~\cite{Keim:2000:DPOVT} visualizations may become satisfying substitutes.

We also plan to provide more advanced network evolution models which support topology evolution and realistic information propagation behaviour.
In such cases, we need at least an extended strategy language to control the simultaneous or alternate (probabilistic) applications of topological modifications or propagation rules.

\subsection*{Acknowledgement}
This work has been partially financed by ANR JCJC EVIDEN 
(2010-JCJC-0201-01) grant. We would also like to thank the 
anonymous reviewers for their detailed comments and helpful suggestions.

\bibliographystyle{eptcs}
\bibliography{references}
\end{document}